\documentclass[prd,a4paper,nofootinbib,
showpacs, 
]{revtex4}
\usepackage{graphicx}
\usepackage{amsfonts}
\usepackage{amssymb}
\usepackage{amsmath}
\usepackage{hyperref}
\usepackage{enumerate}
\usepackage{color}
\def\eq#1{{eq.~(\ref{#1})}}

\def\hhref#1{\href{http://arxiv.org/abs/#1}{#1}} 

\definecolor{oucrimsonred}{rgb}{0.6, 0.0, 0.0}
\definecolor{persianblue}{rgb}{0.11, 0.22, 0.73}
\definecolor{forestgreen}{rgb}{0.13,0.35,0.13}
 \hypersetup{colorlinks, citecolor=oucrimsonred, linkcolor=persianblue, urlcolor=oucrimsonred}
\newcommand{\be}{\begin{equation}}
\newcommand{\ee}{\end{equation}}
\newcommand{\bea}{\begin{eqnarray}}
\newcommand{\eea}{\end{eqnarray}}

\begin{document}
\title[]{Naturalness redux
}
\date{\today}
\author{Marco Fabbrichesi}
\email{marco@sissa.it}
\affiliation{INFN, Sezione di Trieste, via Valerio 2, I-34137 Trieste, ITALY}
\author{Alfredo Urbano}
\email{alfredo.urbano@sissa.it}
\affiliation{SISSA, via Bonomea 265, I-34136 Trieste, ITALY}
\affiliation{INFN, Sezione di Trieste, via Valerio 2, I-34137 Trieste, ITALY}
\begin{abstract}
\noindent  
The idea of naturalness, as originally conceived, refers only to the finite renormalization of the Higgs boson mass induced by the introduction of heavier states. 
In this respect, naturalness is still a useful heuristic principle in model building beyond the standard model whenever 
new massive states are coupled to the Higgs field. The most compelling case is provided by the generation of neutrino masses. In this paper we confront this problem from a new perspective. The right-handed sector responsible for the seesaw mechanism---which introduces a large energy threshold above the electroweak scale---is made supersymmetric  to comply with naturalness while the standard model is  left unchanged and  non-supersymmetric.  Cancellations necessary to the naturalness requirement break down  only at two loops,  thus offering the possibility to increase the right-handed neutrino mass scale up to one order of magnitude above the usual values allowed by naturalness. If also the weak boson sector of the standard model is made supersymmetric, cancellations break down at three loops and the scale of new physics can be further raised. In the type-I seesaw, this  implementation provides  right-handed  neutrino masses  that are natural and at the same time large enough to give rise to   baryogenesis (via leptogenesis). The model contains a dark matter candidate and  distinctive new physics in the leptonic sector.
\end{abstract}

\maketitle

\section{Naturalness reconsidered}
\label{sec:mot}

The idea of naturalness~\cite{nat} is not so much about our distaste for fine-tuning dimensional quantities as about the decoupling of physics at different scales. What is felt to be unnatural is the cancellation among counter-terms originating at different mass scales because it jeopardises the possibility of defining a low-energy theory independently of the details of the higher scale physics. In the  language of effective field theory, the naturalness requirement implies that no large threshold corrections be present. A problem of naturalness only arises if 
\begin{itemize}
\item[1.] There are (at least) two scales. The standard model (SM) taken by itself  is  natural. It is only when new physics is added at a higher scale that the problem arises. It consists in  an unnaturally  fine-tuned cancellation between the bare mass and the quantum corrections in the definition
\be
m_H^2 = m_0^2 - \delta m_H^2\, .
\ee
The correction is proportional to the mass of the new physics states and the coupling of these to the Higgs boson; 
\item[2.] We can compute the effect of the new physics threshold. What cannot be calculated cannot give rise to a naturalness problem because it goes in the unknown (and divergent) part that is present in the quantum corrections. A case in point is gravitational physics: as long as we do not know the theory of quantum gravity, we cannot compute its contribution to the Higgs boson mass and therefore it belongs to the non-computable part of the integral---the part that in dimensional regularization  goes into the residue of the pole and about which we do not worry.  The same holds for the strongly interacting regime at the  Landau pole of the $U(1)_Y$ gauge  interaction. 
\end{itemize}

Even though naturalness as  defined here is the mere restating of well-understood principles~\cite{natgut}, the gist of it is often missed in many approaches (and even in some textbooks) because  the renormalization of the Higgs boson mass is expressed in terms of a momentum-dependent regularization where  integrating over the degrees of freedom of the low-energy theory (the usual loop integral of the quantum corrections) is entangled with the integrating out of the high-scale modes producing the thresholds (in the running of the effective low-energy parameters)~\cite{Rothstein:2003mp}.
 The use of a cutoff regulator---when discussing naturalness, as a proxy for the masses at the higher physical scale---ultimately  leads to incorrect results.  Arguments based on generic  quadratically (and cutoff dependent) divergent  terms are misleading and suggest the mistaken conclusion that radiative corrections from loops of SM particles  give rise to a naturalness problem. These corrections are only of the order of the SM particle masses and therefore included in the physical value of the Higgs boson mass without any fine tuning. In particular,  there is nothing special in the large coupling of the top quark to the Higgs boson and loops of top quarks are not unnaturally large and need not  be compensated by new states. 


The  requirement of naturalness has provided in the past and still  provides today, once properly understood,  a  rather valuable heuristic principle in model building~\cite{natural}.  It tells us that whatever new physics we wish to include  above the SM, to be natural, it must enter in a way that does no introduce large corrections to the Higgs boson mass that would make the electroweak (EW) vacuum unstable.  
On  general grounds, whenever a threshold with particles of mass $M$, coupled to the Higgs field with strength  $\texttt{g}$ is present, 
quantum corrections generate a contribution
\begin{equation}\label{eq:QuadraticCorrections}
\delta m_{H}^2\approx \frac{\texttt{g}^2}{16\pi^2} M^2\, ,
\end{equation} 
to the Higgs boson mass.
It is possible to control the correction in eq.~(\ref{eq:QuadraticCorrections}) in various way. 
In supersymmetry (SUSY), it vanishes because of the presence of bosonic and fermionic states with the same mass~\cite{SUSYcanc}. 
In this paper we confront the problem with the same supersymmetric cancellation mechanism in mind
 but introducing a crucial difference with respect to the usual approach:  it is not the SM that needs to be super-symmetrized but only the new physics sector. On the contrary, the SM breaks SUSY of the new physics explicitly. The scale of SUSY breaking is the EW scale.

If only the new physics sector is made supersymmetric, cancellations  hold at one-loop level.
 They can be preserved to the next loop order by extending SUSY. In such a class of telescoping models, the large corrections are pushed to higher loop levels by  extending SUSY to more and more  sectors of the SM, and then to all loops by re-introducing the fully supersymmetric SM. In this very conservative and pragmatic approach, SUSY is not seen as a fundamental symmetry---its status resembling  that  of custodial symmetry in the SM. The extension of it---and the amount of superpartners---is  determined by how severe the hierarchy problem is and how large a mass we want for the new physics states.

We implement this program in the case of the seesaw mechanism for neutrino masses~\cite{seesaw} as a well-motivated instance of new physics with an higher mass scale above the EW scale.\footnote{In~\cite{Heikinheimo:2013xua} a similar approach was proposed to solve the dark matter (DM) problem.} The  problem in this example is the order of magnitude between the highest mass of the sterile neutrinos that is natural and the lowest possible value for this mass for a viable baryogenesis to take place by means of leptogenesis.


\section{The model}
\label{sec:model}

In this section we illustrate the idea outlined in the introduction with a specific example, namely the generation of neutrino masses via the type-I seesaw mechanism. 
First, in section~\ref{sec:OneGeneration}, we introduce a one-generation toy model which serves as template for  the   more realistic implementation  introduced in section~\ref{sec:ThreeGeneration}. Finally, in section~\ref{sec:SUSYbreaking}, we discuss the two-loops SUSY breaking effects generating  corrections on the Higgs boson mass proportional to the large scale. 

\subsection{From a toy model with one generation...}\label{sec:OneGeneration}

Let us consider first   the (unrealistic) case of a single sterile neutrino with mass $M$ much larger than  the SM scale---which we take to be $v= 264$ GeV, the EW symmetry breaking scale. To protect the EW scale against quantum corrections, this state is introduced in a supersymmetric manner by means of the following  chiral supermultiplet\footnote{Throughout this paper we adopt the notation of ref.~\cite{Martin:1997ns}.}
\be\label{eq:NP}
\Phi_{\rm NP} = \phi + \theta\cdot N + \theta^2 F\,,
\ee
which is a singlet under the SM gauge groups and is  made of the fields encoding the new physics:
we identify the bosonic component $\phi$ with an inert (complex) scalar singlet, the fermionic component $N$ with the Weyl component of a Majorana sterile neutrino.\footnote{Since we are working with left-handed chiral superfields, $N$ should be interpreted as the charge conjugate of a right-handed sterile neutrino.}
Furthermore,  we assign to $\Phi_{\rm NP}$ a $r_{\Phi_{\rm NP}} = 1$ charge under a $U(1)_R$ symmetry
to prevent the presence of linear or trilinear terms in $\Phi_{\rm NP}$ in the superpotential.

The supersymmetric  sector is described by the usual  Wess-Zumino Lagrangian~\cite{Wess:1973kz} density
\be
\mathcal{L}_{\rm NP} = \int d^2 \theta d^2 \bar \theta \:  \Phi_{\rm NP}^\dag \Phi_{\rm NP}  + 
\left[ \int d^2 \theta \: {\cal W}(\Phi_{\rm NP}) + h.c. \right]\,,
\ee
where the superpotential is given by
\be\label{pot}
{\cal W}(\Phi_{\rm NP}) =   \frac{M}{2} \Phi_{\rm NP}^2\,.
\ee
The mass $M$ is the scale of new physics. 

The equation of motion for the auxiliary field is
\be
F = - \left. \frac{d {\cal W^\dag}}{d \Phi_{\rm NP}^\dag} \right|_{\Phi_{\rm NP}^\dag=\phi^*} =  -  M\phi^*\,.
\ee

At this stage the SM is described by the usual non-supersymmetric Lagrangian. In order to couple it to the new physics sector we
identify a supersymmetric chiral superfield inside the SM states, namely
\be\label{eq:L}
\Phi_1=  \tilde H + \theta\cdot L + \theta^2 F_1\,, 
\ee
where $\Phi_1$ is a $SU(2)_L$ doublet made of the SM Higgs boson $\tilde H = i\sigma_2 H^*$ and the lepton doublet field (the left-handed electron and the corresponding neutrino). There is no contribution from $\Phi_1$ to the superpotential   because of gauge invariance. The multiplet components have their SM masses, the splitting of which is of the order of the EW scale and sets the SUSY breaking scale provided by the SM Lagrangian. 

To protect the EW scale, the coupling must be  such as  that no corrections $\mathcal{O}(M^2)$ are introduced.  
It is not possible to couple the new physics chiral superfield $\Phi_{\rm NP}$ and the SM chiral superfield $\Phi_1$  in a supersymmetric manner because of the hypercharge assignments. There are two solutions to this problem and correspondingly two possible models. Either one works with just $\Phi_1$ and introduces the interaction in the K\"ahler potential, or one introduces a new scalar multiplet $\Phi_2$ with the opposite hypercharge---as it is done in the Minimal Supersymmetric Standard Model (MSSM). Both models produce the same Yukawa coupling between the sterile and the SM neutrino. They can be distinguished because they come with their own characteristic set of new states. 
To remain with the simplest case  (that is, without the necessity of introducing higher-dimensional operators in the K\"ahler potential), we explore in this paper the second possibility.
Therefore,  the chiral superfields are now three:
 \bea
 \Phi_{1}^{\alpha} & =&  H^u_{\alpha} + \theta\cdot L_{\alpha}  + \theta^2 F^u_{\alpha}~,\label{eq:Chiral1}\\
   \Phi_{2}^{\alpha} &=& H^d_{\alpha} + \theta\cdot \tilde{h}^d_{\alpha} + \theta^2 F^d_{\alpha}~,\label{eq:Chiral2}\\
   \Phi_{\rm NP} &=& \phi + \theta\cdot N + \theta^2 F\,,\label{eq:ChiralNP}
 \eea
 where the index $\alpha$ refers to the $SU(2)_L$  gauge group. We assign the following $U(1)_Y$, $SU(2)_L$, and $U(1)_R$ charges:
\bea
\Phi_1: &  \quad (Y=-1,~T_3=1/2,~r_{\Phi_1} = 0)~,\label{eq:charges1} \\
  \Phi_2: & \quad (Y=+1,~T_3=1/2,~r_{\Phi_2}=1)~,\label{eq:charges2} \\
   \Phi_{\rm NP}: &\quad (Y=0,~T_3=0,~r_{\Phi_{\rm NP}}=1)\,.\label{eq:charges3}
\eea 
In eq.~(\ref{eq:Chiral2}) $\tilde{h}^d$ represents a $SU(2)_L$ doublet of left-handed Weyl fermions, $\tilde{h}^d = (\tilde{h}^+_d,\tilde{h}^0_d)_{L}^{\rm T}$
 while in eqs.~(\ref{eq:Chiral1})--(\ref{eq:Chiral2}) we adopt, as customary in models with two Higgs doublets (see section~\ref{sec:Higgses}), the following parametrization (before  EW symmetry breaking) for the four neutral and two charged degrees of freedom:
 \begin{eqnarray}
 H^u &=&
\left(
\begin{array}{c}
\frac{1}{\sqrt{2}} (H^0\cos{\alpha} - h_0\sin\alpha +i A^0\sin\beta -iG^0\cos\beta)  \\
  H^-\sin\beta - G^-\cos\beta 
\end{array}
\right)\,,\label{eq:Hup}\\
 H^d &=&
\left(
\begin{array}{c}
 H^+\cos\beta + G^+\sin\beta  \\
\frac{1}{\sqrt{2}} (H^0\sin\alpha + h_0\cos\alpha +i A^0\cos\beta + iG^0\sin\beta)
\end{array}
\right)\,,\label{eq:Hdown}
 \end{eqnarray}
 where $\alpha$, $\beta$ are two mixing angles and where we identify the neutral $h_0$ component with the physical Higgs of the SM.

 The interactions among the three chiral superfields $\Phi_1$, $\Phi_2$ and $\Phi_{\rm NP}$ in this model can be written in a supersymmetric  manner, and the superpotential is given by
 \be
 {\cal W}(\Phi_i)  =  \eta \, \epsilon_{\alpha \beta} \Phi_1^\alpha \Phi_2^\beta \Phi_{\rm NP} + \frac{M}{2} \Phi_{\rm NP}^2\,, \label{pot}
 \ee
 where $\eta$ is a generic coupling, while $\epsilon \equiv i\sigma_2$ indicates the $SU(2)_L$ invariant product (for simplicity, we take $\eta$ and $M$ reals). 
 All other terms are forbidden by gauge or $U(1)_R$-symmetry because of our assignment in eqs.~(\ref{eq:charges1})--(\ref{eq:charges3}).
 There is a gauge anomaly that can be compensated  by introducing an extra state with the appropriate hypercharge and mass
 $\mathcal{O}(m_h)$---so as not to introduce new high-scale masses. 
 The presence of this additional Higgsino-like state has important phenomenological consequences that will be discussed in the next section.
 
 By adding canonical kinetic terms for $\Phi_{1,2}$, the model is described by the following Lagrangian
 \begin{eqnarray}\label{lag}
 \mathcal{L} &=& (\partial_{\mu}\phi)^*(\partial^{\mu}\phi) + \bar{N}i\bar{\sigma}^{\mu}(\partial_{\mu}N) + \bar{L}i\bar{\sigma}^{\mu}(\partial_{\mu}L) +  \bar{\tilde{h}}^d i\bar{\sigma}^{\mu}(\partial_{\mu}\tilde{h}^d)
 + (\partial_{\mu}H_{\alpha}^{d})^*(\partial^\mu H^d_{\alpha}) + (\partial_{\mu}H^u_{\alpha})^*(\partial^\mu H^u_{\alpha})\nonumber \\
 &+& F^*F+(F_{\alpha}^u)^*F_{\alpha}^u+(F_{\alpha}^d)^*F_{\alpha}^d + M\phi F + M\phi^* F^* -\frac{M}{2}N\cdot N -\frac{M}{2}\bar{N}\cdot \bar{N}\nonumber \\
 &+&\eta\epsilon_{\alpha\beta}
 \left(
 H^u_{\alpha}H^d_{\beta}F + H_{\beta}^d\phi F^u_{\alpha} + \phi H^u_{\alpha}F^d_{\beta} - L_{\alpha}\cdot \tilde{h}^d_{\beta}\phi  -  \tilde{h}^d_{\beta}\cdot N H^u_{\alpha} 
 - L_{\alpha} \cdot N H^d_{\beta} + h.c.
 \right)\,.
 \end{eqnarray}
The equations of motion of the auxiliary fields are now
\be
F^u_\alpha = - \eta \, \epsilon_{\alpha \beta} H^{d\,*}_\beta \phi^*~,~~~F^d_\beta = - \eta \, \epsilon_{\alpha \beta} H^{u\,*}_\alpha \phi^*~,~~~
F = -   \eta \, \epsilon_{\alpha \beta}  H^{u\,*}_\alpha H^{d\,*}_\beta -  M \phi^*\,,
\ee
and replacing them in \eq{lag} gives rise to the  following kinetic and interaction Lagrangians 
 \begin{eqnarray}
 \mathcal{L_{\rm kin}} &=& (\partial_{\mu}\phi)^*(\partial^{\mu}\phi) - M^2|\phi|^2 + \bar{N}i\bar{\sigma}^{\mu}(\partial_{\mu}N) 
 -\frac{M}{2}\left(N\cdot N + \bar{N}\cdot \bar{N}\right)\label{eq:Kin} \\
 &+& \bar{L}i\bar{\sigma}^{\mu}(\partial_{\mu}L) + \bar{\tilde{h}}^d i\bar{\sigma}^{\mu}(\partial_{\mu}\tilde{h}^d)
 + (\partial_{\mu}H_{\alpha}^{d})^*(\partial^\mu H^d_{\alpha}) + (\partial_{\mu}H^u_{\alpha})^*(\partial^\mu H^u_{\alpha})~, \nonumber \\
 \mathcal{L_{\rm int}} &=& -\eta^2(\epsilon_{\alpha\beta}H^{u}_{\alpha}H^d_{\beta})(\epsilon_{\alpha'\beta'}H_{\alpha'}^{u\,*}H_{\beta'}^{d\,*})
 -\eta^2|\phi|^2H_{\alpha}^dH_{\alpha}^{d\,*}-\eta^2|\phi|^2H_{\alpha}^uH_{\alpha}^{u\,*}\label{eq:int}\\
 &-&\eta \epsilon_{\alpha\beta} \left(
 M\phi^*H^u_{\alpha}H^{d}_{\beta}
 +L_{\alpha}\cdot \tilde{h}^d_{\beta}\phi
 +\tilde{h}^d_{\beta}\cdot N H^u_{\alpha} + L_{\alpha}\cdot N H^d_{\beta}
 +h.c.
 \right)\,.\nonumber
 \end{eqnarray}

The Yukawa coupling $\epsilon_{\alpha\beta}L_{\alpha}\cdot N H^d_{\beta} + h.c.$---together with the Majorana 
 mass term for $N$---generates the usual type-I seesaw mechanism. In eq.~(\ref{eq:int}) the  quartic term with both scalar fields and the mixed term proportional to $M$ 
guarantee, at the one-loop level, the typical cancellation characteristic of the unbroken Wess-Zumino model: the sum of the corrections to the Higgs boson mass vanishes
 as long as the components of the new physics multiplet have the same mass. 
\begin{figure}[!htb!]
  \includegraphics[width=0.6\textwidth]{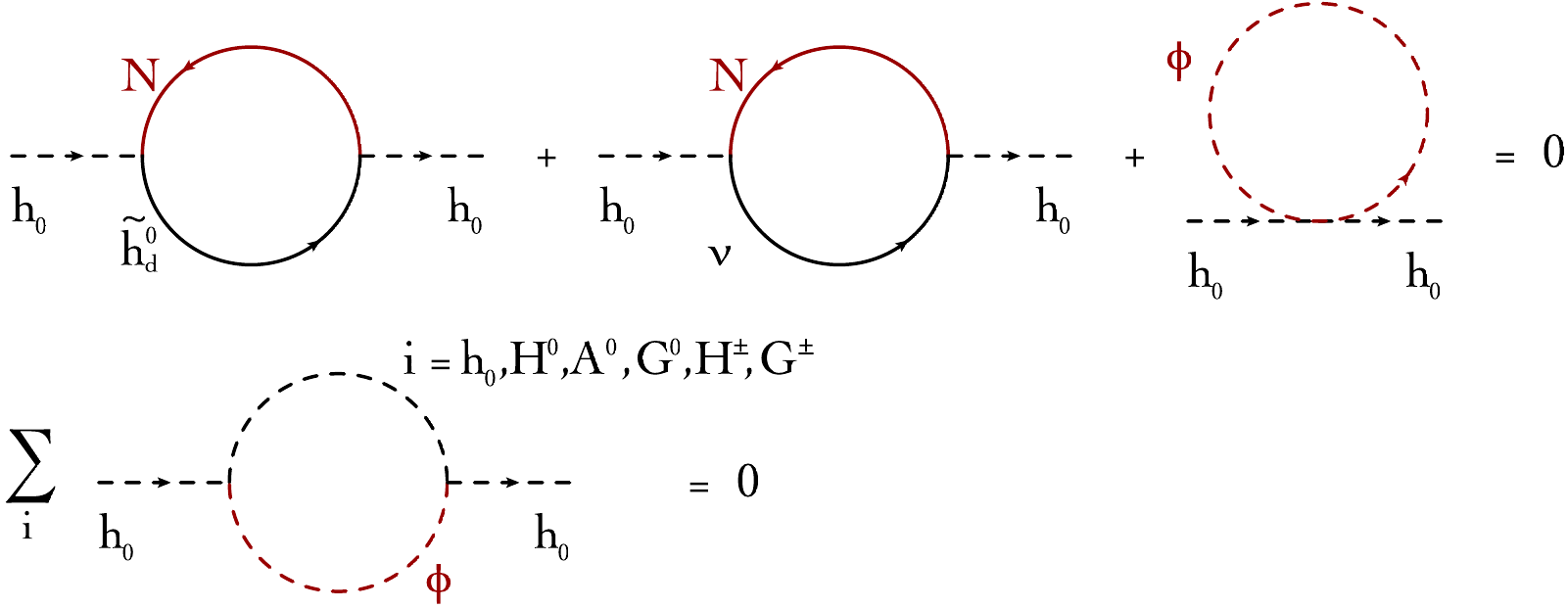}
 \caption{\textit{
One-loop cancellation among different contributions to the Higgs boson mass term in the setup outlined in section~\ref{sec:OneGeneration}. 
We mark in red the heavy states with mass $M$. 
 }}\label{fig:OneLoopCancellation}
\end{figure}
For example the correction to the Higgs boson mass coming from the loop with the SM lepton and the sterile neutrino is cancelled by the sum of the diagram with the scalar singlet in the loop and the one with the scalar singlet and the fermionic neutral component of the chiral superfield $\Phi_2$. We summarize in fig.~\ref{fig:OneLoopCancellation}, 
from a diagrammatic point of view, the cancellation mechanism at the one-loop level.
Notice that
at this stage we are working in the unbroken EW phase since soft terms (like the masses of SM particles) only introduce natural corrections proportional to the EW scale.

Let us close this section with one final technical remark that will be useful in section~\ref{sec:np}. Starting from the Lagriangian
in eqs.~(\ref{eq:Kin})--(\ref{eq:int}), it is straightforward to integrate out the heavy fields $N$ and $\phi$. We find
\begin{eqnarray}\label{eq:EffL}
 \mathcal{L}_{\rm eff} &=& \bar{L}i\bar{\sigma}^{\mu}(\partial_{\mu}L) + \bar{\tilde{h}}^d i\bar{\sigma}^{\mu}(\partial_{\mu}\tilde{h}^d)
 + (\partial_{\mu}H_{\alpha}^{d})^*(\partial^\mu H^d_{\alpha}) + (\partial_{\mu}H^u_{\alpha})^*(\partial^\mu H^u_{\alpha})\\
 &+& \frac{\eta^2}{M}
  \epsilon_{\alpha\beta}\epsilon_{\alpha'\beta'}\left[
L_{\alpha}\cdot \tilde{h}^d_{\beta}
  H_{\alpha'}^uH_{\beta'}^{d}
  +\frac{1}{2}L_{\alpha}\cdot L_{\alpha'}H_{\beta}^dH^d_{\beta'}
  +\frac{1}{2}\tilde{h}^d_{\alpha}\cdot \tilde{h}^d_{\alpha'}H_{\beta}^uH^u_{\beta'}
  -L_{\alpha}\cdot \tilde{h}^d_{\alpha'}H_{\beta}^dH_{\beta'}^u + h.c.
 \right]+\mathcal{O}(M^{-2})\,,\nonumber
\end{eqnarray}
where the first term (last three terms) in the square bracket is (are) generated by integrating out $\phi$ ($N$). The dim-$5$ operator $\epsilon_{\alpha\beta}\epsilon_{\alpha'\beta'}L_{\alpha}\cdot L_{\alpha'}H_{\beta}^dH^d_{\beta'}$ is the familiar (in spinor notation) Weinberg operator for neutrino masses. From eq.~(\ref{eq:EffL}) is clear that all the effects  due to the heavy chiral superfield $\Phi_{\rm NP}$ decouple for large values of $M$.
Finally notice that, as customary in SUSY, all the effective operators in eq.~(\ref{eq:EffL}) are controlled by the same coupling constant, namely the Yukawa coupling $\eta$ of the right-handed neutrino.
 
\subsection{...to a more realistic implementation}\label{sec:ThreeGeneration}

A realistic implementation requires more than one generation of neutrinos.  
Every supersymmetrized leptonic multiplet brings into the model an extra gauge doublet of scalar leptons which plays the role of the Higgs boson doublet in the first generation multiplet.
The chiral superfields in \eq{eq:Chiral1} and \eq{eq:ChiralNP} are now replaced by
\be\label{eq:FieldContent}
\Phi_1^a = \tilde H^a  + \theta\cdot L^a  + \theta^2 F_1^a \, \quad \mbox{and} \quad 
\Phi_{\rm NP}^a = \phi^a  + \theta\cdot N^a  + \theta^2 F^a\,,
\ee
where $\tilde{H}^1$ plays the role of  the Higgs boson doublet $H^u$ in section~\ref{sec:OneGeneration}, while $\tilde{H}^{2,3} \equiv \tilde L^{2,3}$ are scalar leptons doublets. The multiplet $\Phi_2$ is left unchanged.
There are three kinds of sterile neutrinos and, correspondingly, three kinds of additional heavy scalar singlets.\footnote{A simpler setup would be with just two kinds of sterile neutrinos~\cite{Frampton:2002qc}.} 

The  superpotential  becomes
\be
{\cal W}(\Phi_i) =   \hat \eta_{ab} \epsilon_{\alpha \beta} \Phi_1^{\alpha, a} \Phi_2^{\beta} \Phi_{\rm NP}^b + \frac{1}{2} \hat M_{ab} \Phi_{\rm NP}^a \Phi_{\rm NP}^b\,,
\ee
and the relevant interaction Lagrangian is therefore
\begin{equation}\label{eq:SUSYsector}
\mathcal{L} \ni -\epsilon_{\alpha\beta} \left(\hat \eta_{ab}L_{\alpha}^a\cdot N^b H_{\beta}^d 
+\hat \eta_{ab}L_{\alpha}^a\cdot \tilde{h}^d_{\beta}\phi^b
+\hat \eta_{1b}\tilde{h}^d_{\beta}\cdot N^b H^u_{\alpha}
+\hat \eta_{2b}\tilde{h}^d_{\beta}\cdot N^b \tilde{L}^2 
 +\hat \eta_{3b}\tilde{h}^d_{\beta}\cdot N^b \tilde{L}^3  + h.c.
\right)\,.
\end{equation}

The first term in \eq{eq:SUSYsector} gives the seesaw Yukawa coupling between sterile neutrinos and leptons, 
while the other terms give new physics beyond the SM, namely the Yukawa terms for the two scalar leptons interacting with sterile neutrinos and Higgsinos. 
These interactions always involve one heavy state---either $N$ or $\phi$---and are not relevant for low-energy phenomenology. 
We discuss some aspects of this phenomenology in section~\ref{sec:np}.

\subsection{SUSY breaking at two and three loops}\label{sec:SUSYbreaking}

\begin{figure}[!htb!]
  \includegraphics[width=0.6\textwidth]{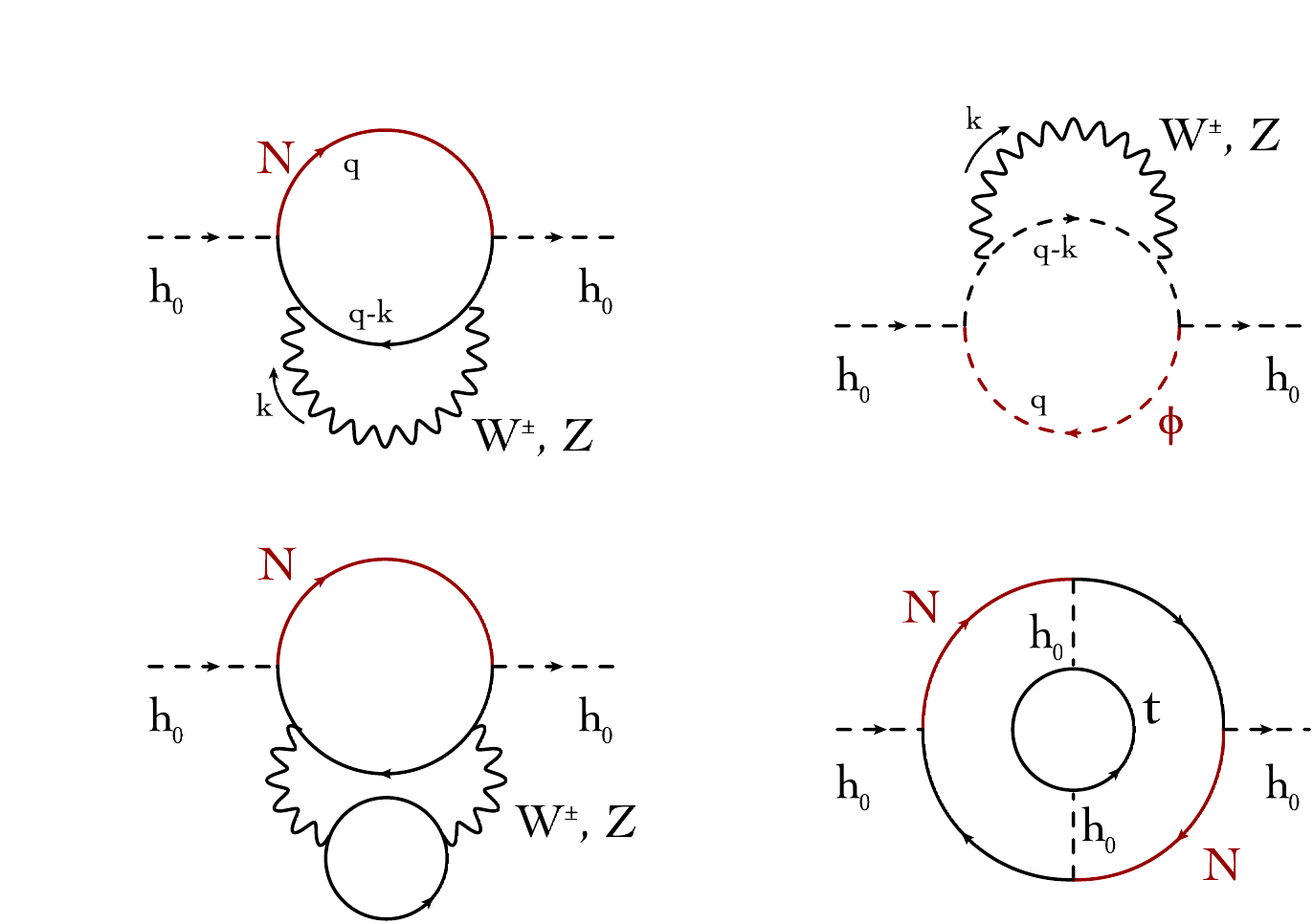}
 \caption{\textit{
Upper panel: representative two-loop corrections to the Higgs boson mass responsible for the quadratic correction proportional to $M^2$ in eq.~(\ref{eq:M2}).
The heavy states with mass $M$ are marked in red. Lower panel: representative three-loop corrections relevant if SUSY is extended also to the EW gauge sector of the SM (see text for details).
 }}\label{fig:TwoLoopNonCancellation}
\end{figure}

The SM is not supersymmetric. It couples to the supersymmetric sector via the EW gauge interactions which therefore break explicitly SUSY at the two-loop level. 
This can be understood by looking at diagrams in which a loop of gauge particles is added to the basic one-loop diagrams of the previous section. Representative diagrams of this kind are shown in the upper panel of fig.~\ref{fig:TwoLoopNonCancellation}. 

Since the quantity we are interested in---the two-point function of the Higgs boson with zero external momentum in the massless EW theory---is gauge invariant, the computation can be performed in the Landau gauge, where the only relevant diagrams are those in the upper panel of  fig.~\ref{fig:TwoLoopNonCancellation}.
 Neglecting all masses at the EW scale we find, considering for simplicity the model with one generation
\bea\label{eq:Master2Loop}
\delta m_H^2 & =&  -\frac{\eta^2g^2}{2}\left(1 + \frac{1}{2\cos^2\theta_{\rm W}}\right)\int\frac{d^4q}{(2\pi)^4}\frac{d^4k}{(2\pi)^4}
\frac{[q^2-q\cdot k(3-2q\cdot k/k^2)]}{(q - k)^2q^2(q^2 - M^2)k^2}\nonumber \\
& & +
\eta^2g^2M^2\cos^2\alpha\int\frac{d^4q}{(2\pi)^4}\frac{d^4k}{(2\pi)^4}\frac{[q^2 - (q\cdot k)/k^2]}{q^4 k^2(q - k)^2(q^2 - M^2)}\,,
\eea
where $g$ is the $SU(2)_L$ gauge coupling and $\theta_W$ the Weinberg angle. In eq.~(\ref{eq:Master2Loop}) the first (second) integral comes from the first (second) diagram in the upper panel of fig.~\ref{fig:TwoLoopNonCancellation}, where the sum over all the particles exchanged in the loop is understood. 

 The two-loop integrals can be computed using dimensional regularization~\cite{Jurcisinova:2010zz}; we find
\begin{eqnarray}
\delta m_H^2 & =& -\frac{\eta^2g^2}{2}\left(1 + \frac{1}{2\cos^2\theta_{\rm W}}\right)
\left[
\frac{M^2}{(4\pi)^4}\left(\frac{4\pi\mu^2}{M^2}\right)^{2\epsilon}\Gamma(2\epsilon-1)\int_0^1dx_1dx_2dx_3 \frac{\delta(x_1 +x_2 +x_3 -1)x_3^{d-3}}{(x_1x_2 + x_1x_3 +x_2x_3)^{d/2}}
\right] \\
&& -\eta^2g^2M^2\cos^2\alpha
\left[
\frac{1}{(4\pi)^4}\left(\frac{4\pi\mu^2}{M^2}\right)^{2\epsilon}\Gamma(2\epsilon)\int_0^1
dx_1dx_2dx_3dx_4 \frac{\delta(x_1 +x_2 + x_3 + x_4 -1)x_4^{d-4}}{(x_1x_2 + x_1x_3 +x_2x_3 + x_2x_4 + x_3x_4)^{d/2}}
\right]~,\nonumber
\end{eqnarray}
in $d = 4-2\epsilon$ dimensions. The integrals over the Feynman parameters  can be evaluated by means of the Cheng-Wu theorem~\cite{Smirnov:2012gma} and, in the $\epsilon \to 0$ limit, 
we find the following exact result
\begin{equation}
\delta m_H^2 = \frac{\alpha_{\rm W}}{4\pi}\frac{\eta^2 M^2}{16\pi^2}
\left(
\frac{1}{2}+\frac{1}{4\cos^2\theta_{\rm W}} - \cos^2\alpha
\right)
\left[
\frac{1}{2\epsilon^2} +\frac{1}{\epsilon}
\left(
\frac{3}{2}-\gamma_{\rm E} +\ln\frac{4\pi\mu^2}{M^2}
\right) + \mathcal{C}
\right]~,
\end{equation}
where $\gamma_{\rm E}$ is the Euler-Mascheroni constant and 
\begin{equation}
\mathcal{C} \equiv \frac{7}{2} -3\gamma_{\rm E} +\gamma_{\rm E}^2 + \frac{\pi^2}{4}
+3\ln\frac{4\pi\mu^2}{M^2} - 2\gamma_{\rm E}\ln\frac{4\pi\mu^2}{M^2} + \ln^2\frac{4\pi\mu^2}{M^2}~.
\end{equation}
Disregarding the UV divergent part, and taking the renormalization scale $4\pi\mu^2 = M^2$, we find 
\be\label{eq:M2}
\delta m_H^2 = \frac{\alpha_{\rm W}}{4\pi}\frac{\eta^2 M^2}{16\pi^2}\left(
\frac{1}{2}+\frac{1}{4\cos^2\theta_{\rm W}} - \cos^2\alpha
\right)\left(
\frac{7}{2} -3\gamma_{\rm E} +\gamma_{\rm E}^2 + \frac{\pi^2}{4}
\right)
\simeq \frac{\alpha_{\rm W}}{4\pi}\frac{\eta^2 M^2}{16\pi^2}~,
\ee
where we approximate the numerical combination in parenthesis with a $\mathcal{O}(1)$ factor. 
The generalization of this formula to the case of three generation would just introduce 
a more complicated factor with the general structure 
${\rm Tr}(\hat \eta^{\rm T}\hat M^2\hat \eta)$ (assuming, without loss of generality, a diagonal $\hat M$). 
Since we are interested more in an order-of-magnitude estimate 
rather than in a precise numerical analysis,
we will employ eq.~(\ref{eq:M2}) for the purposes of our discussion.

Before illustrating the implication of eq.~(\ref{eq:M2}), let us notice that the cancellation observed at one loop (as discussed in section~\ref{sec:OneGeneration} and fig.~\ref{fig:OneLoopCancellation}) can be further extended at two loops by super-symmetrizing the $U(1)_Y\otimes SU(2)_L$ gauge sector of the SM.
In this case  corrections to  $\delta m_H^2$ proportional to  $M^2$ arise only at three loops via exchange of SM quarks and leptons, as shown  in the bottom panel of fig.~\ref{fig:TwoLoopNonCancellation}. 
\begin{figure}[!htb!]
  \includegraphics[width=0.5\textwidth]{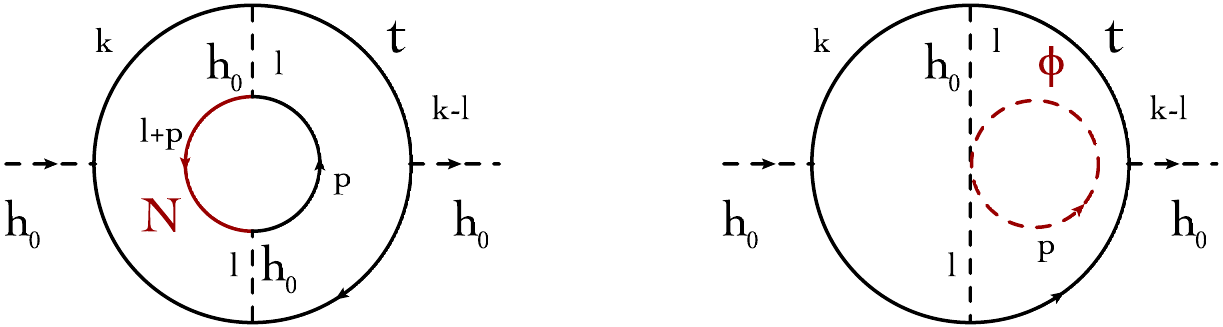}
 \caption{\textit{
 Three loops double-bubble diagrams with the top on the external bubble.
 }}\label{fig:TwoLoopMiracle}
\end{figure}
Among these three-loop dcontributions, the double-bubble diagrams in fig.~\ref{fig:TwoLoopMiracle} deserve special attention. 
These diagrams are proportional to the fourth power of the Yukawa top coupling $y_t$, and---if proportional to $M^2$---they  
would
generate a correction to $\delta m_H^2$ comparable in magnitude with the two-loop result in eq.~(\ref{eq:M2}).
We find 
\begin{eqnarray}\label{eq:ThreeLoopMiracle}
\delta m_H^2 =&& -i\eta^2y_t^4\int\frac{d^4k}{(2\pi)^4}\frac{d^4l}{(2\pi)^4}\frac{d^4p}{(2\pi)^4}\times \nonumber \\
&&\left\{
-\frac{1}{k^2(k-l)^2l^4[(p + l)^2 - M^2]}
+
\frac{1}{k^2(k-l)^2l^4(p^2 - M^2)} -\frac{p\cdot k}{k^2(k-l)^2l^4p^2[(p+l)^2 - M^2]}\right\}~.
\end{eqnarray}
The first two integrals in eq.~(\ref{eq:ThreeLoopMiracle})---which correspond, respectively, to the scalar part of the 
two loops over $N$ and $\phi$ in the internal bubbles in fig.~\ref{fig:TwoLoopMiracle}---generate, if computed individually, a correction proportional to $M^2$. However their sum, protected by SUSY, vanishes. The third integral in eq.~(\ref{eq:ThreeLoopMiracle})---which corresponds to the tensor part of the loop over $N$ 
in the left diagram in fig.~\ref{fig:TwoLoopMiracle}---gives only a correction 
proportional to the external momentum (that is, a  contribution to the wave-function renormalization). 
Therefore, the class of double-bubble integrals in fig.~\ref{fig:TwoLoopMiracle} cannot generate large $M^2$ corrections
if SUSY protects the Higgs two-point function in the internal bubble.


\subsection{On the largest mass scale allowed by naturalness}\label{eq:NeutrinoMasses}

As first discussed in~\cite{Vissani:1997ys}, the type-I seesaw mechanism is natural for sterile neutrino masses up to $M \simeq 10^7$ GeV.\footnote{The case of 
low-scale (resonant) leptogenesis~\cite{Pilaftsis:2003gt} is natural 
because it can be realized with heavy Majorana neutrinos as light as the EW scale (\cite{Pilaftsis:2005rv}; for a
complete analysis including all flavor effects, see e.g.~\cite{Dev:2014laa}).}
Up to this value, the Higgs boson mass can have its physical value without requiring any fine-tuned cancellation.  In fact, the one-loop correction to the Higgs boson mass is given
\be
\delta m_H^2 \simeq  \frac{y_\nu^2}{16 \pi^2}  \, M^2  \,,  \label{1}
\ee
where $y_\nu$ stands for the Yukawa couplings of the sterile neutrinos and leptons. These  couplings are fixed by the neutrino mass value and given by
\be
y_\nu^2 = M m_{\nu}/v^2 \,.
\ee
By taking $m_\nu =0.05$ eV, and stretching  the naturalness condition to its upper bound by taking $\delta m_H^2 \lesssim (1\,{\rm TeV})^2$, we find
\be
M \lesssim  10^8 \; \mbox{GeV} \, .
\ee 
Notice that $M\sim 10^8$ GeV corresponds to a Yukawa coupling $y_\nu \sim 10^{-4}$.
The result of ~\cite{Vissani:1997ys} can be reproduced by taking, more conservatively, $\delta m_H^2 \lesssim v^2$.\footnote{
Notice how, by following the cutoff prescription discussed in the introduction, in this case we would have been misled to consider the loop of the top quark and write
$
\delta m_H^2 \simeq  \lambda_t \Lambda^2  /(16 \pi^2) 
$
and erroneously find 1 TeV as the largest value for having a natural seesaw.}

On the other hand, in order for baryogengesis to proceed via leptogenesis the mass of the sterile neutrinos must 
be of the order of $10^9$ GeV---the specific  number  depending on the assumptions on the initial abundance and the hierarchy in the right-handed neutrino states~\cite{Davidson:2002qv}. This tension  has been confirmed recently in \cite{Clarke:2015gwa} for the case with three generations of neutrinos.

The model introduced in the previous section  helps in  solving this problem. 
The coupling $\eta$ in eq.~(\ref{eq:int}) can be identified with the Yukawa coupling $y_{\nu}$ previously introduced. 
The cancellation  taking place at the one-loop level pushes the first contribution to the Higgs boson mass  at the two-loop level. This suppression can  be used to rise the scale of the heavy states.  As we have seen in the previous section,  the two-loop corrections gives a shift in the Higgs boson mass similar to that above but rescaled, namely
\be
\delta m_H^2 \simeq  \frac{y_{\nu}^2}{16\pi^2}\frac{\alpha_{\rm W}}{4\pi}M^2~, 
\ee
which is natural  as long as it is of the order of the EW scale and at most 1 TeV.  We now find that the sterile neutrino mass can be as large as
\be
M \simeq  10^9 \; \mbox{GeV} \, ,
\ee 
and therefore close to  the lower bound on baryogenensis---the actual number depending on the details of the implementation.
 
 If the supersymmetric sector of the SM is extended, by telescoping this model, to include also the EW gauge bosons, 
this value can be increased by one order of magnitude to $M \simeq  10^{10}$ GeV and thus  comfortable within the viable range for leptogenesis.\footnote{In order to produce thermally the heavy neutrino states a reheating temperature of the Universe after inflation of $T_{\rm R}\gtrsim M$ is required;
excessively high values of $T_{\rm R}$ typically lead
to an overproduction of gravitinos in the early Universe, in tension with Big Bang Nucleosynthesis (BBN) constraints~\cite{Kawasaki:1994af}.
However, reheating temperatures up to $T_{\rm R} \simeq 10^{10}$ GeV---compatible with a mass scale $M\simeq 10^{9}$---
are not excluded because the gravitino density is still too small to cause disruption of BBN~\cite{Davidson:2002qv}.
Since our SUSY scenario does not descend from a supergravity model, we do not expect in any case to encounter the gravitino problem.} Larger scales can be achieved by extending SUSY to the SM fermions and then to the colored gauge states. In the latter case, we recover the full MSSM and cancellations to all loop orders.
From this perspective, our approach completely overturns the usual conclusion (based on the cutoff regularization) according to which, 
in order to cancel the contribution of the top loop, one is forced to introduce, first and foremost, colored top partners in order  to naturally justify a new physics scale  above the TeV.



\section{New physics}
\label{sec:np}

In this section we discuss some low-energy implication of the setup outlined in section~\ref{sec:model}. 
In particular, in section~\ref{sec:Leptons} we highlight the new interactions characterizing the leptonic sector of the SM Lagrangian, and 
in section~\ref{sec:ExtraScalars} we discuss the role of extra scalar states;  in section~\ref{sec:Higgses}, we briefly mention the properties related to the existence of two Higgs doublets and conclude, in section~\ref{sec:dm}, reviewing possible DM candidates.

The aim of this discussion is not to be exhaustive, since a careful phenomenological analysis is beyond the scope of this paper.  

\subsection{Leptonic sector}\label{sec:Leptons}

At low energy the SM Lagrangian is enlarged by new degrees of freedom originating from the supersymmetric sector, $\mathcal{L}=\mathcal{L}_{\rm SM} + \mathcal{L}_{\rm NP}$.

Focusing on the leptonic sector, in addition to the usual SM Yukawa coupling, the model features 
the Yukawa interaction involving the right-handed neutrinos in eq.~(\ref{eq:SUSYsector}) together with their mass and kinetic term 
\begin{equation}\label{eq:RelevantLeptonSector}
\mathcal{L}_{\rm NP} \supset
\overline{N^a_R}i\gamma^{\mu}\partial_{\mu}N^a_R - \left[ \hat \eta_{ab}\overline{N_R^a}\nu_L^b H_2^d + \frac{\hat M_{ab}}{2}\overline{N_R^a}(N_R^b)^{\rm C} + h.c. \right]~,
\end{equation}
The role of the SM Higgs doublet is played, as already mentioned, by $H^d$. Furthermore, without loss of generality, we work in the basis in which $\hat M$ is diagonal. 
At this stage, lepton number is an accidental symmetry of the SM Lagrangian 
with the usual assignments. As a consequence in eq.~(\ref{eq:RelevantLeptonSector})
lepton number is violated by the Majorana mass term for two units as customary in the type-I seesaw.
Notice that in our model we are not trying to relate the lepton number to the $U(1)_R$ charge of the supersymmetric sector. On the contrary,  lepton number is badly broken 
by the interactions in eq.~(\ref{eq:SUSYsector}). 
The important point to keep in mind is that all the interactions in eq.~(\ref{eq:SUSYsector}) always involve one component of the heavy
chiral superfield $\Phi_{\rm NP}$ and, apart form neutrino masses, they decouple from low-energy phenomenology.
 Since we are dealing with 
 sterile neutrino masses as large as $10^8$ GeV,  in this realization all the effects related to the mixing between light and heavy neutrinos are severely suppressed.

\subsection{Extra scalars}\label{sec:ExtraScalars}

With reference to eq.~(\ref{eq:FieldContent}), the model---considering the realization discussed in this paper---provides 
for extra scalar $SU(2)_L$ doublets with hypercharge $Y=-1$, namely $\tilde{L}^{2,3}$. 
The Lagrangian $L_{\rm NP}$, therefore, will contain their kinetic and bare mass terms 
\begin{equation}\label{eq:LagrangianLtilde}
L_{\rm NP} \supset  (D_{\mu}\tilde{L}^i)^{\dag}(D^{\mu}\tilde{L}^i ) - m_{\tilde{L}^i}^2 (\tilde{L}^i)^{\dag}\tilde{L}^i~,
\end{equation}
where, in a diagonal basis, $i=2,3$. This mass term is protected by the same mechanism protecting that of the Higgs boson.
Additional couplings and mixings, allowed by Lorentz and gauge symmetries, might be present; for the sake of simplicity, we limit our analysis to a minimal set of operator. 
Since $\tilde{L}^{2,3}$ have gauge interactions, they can be {\it i)} constrained using LEP data via direct searches and EW precision observables and {\it ii)} produced directly at the LHC via EW Drell-Yan processes mediated by $s$-channel $Z$ or $W^{\pm}$ exchange. In both cases it is crucial to discuss the mass splitting between the neutral and the charged component of the doublet.

\begin{enumerate}[{}\it i)] 

\item 
LEP searches for both neutral and charged scalar particles place a lower limit on the value of their mass; as a rough estimate, 
we can use a representative $\mathcal{O}(100\,{\rm GeV})$ lower limit  from direct slepton searches~\cite{LEPIISUSY}.
As far as the EW precision observables are concerned, the presence of the scalar doublet $\tilde{L} = (\tilde{L}^0, \tilde{L}^-)^{\rm T}$ (we omit the index $i$ hereafter)
affects the oblique parameters $S$, $T$ and $U$ \cite{Peskin:1990zt}. These parameters enter in the vacuum polarization amplitudes of the EW gauge bosons, and are severely constrained by LEP-I and LEP-II results \cite{Barbieri:2003pr}. 
For the scalar doublet $\tilde{L}$ we find the contributions~\cite{Blum:2011fa}
\begin{eqnarray}
\alpha S_{\tilde{L}} &=& \frac{1}{12\pi}\ln\frac{m_{\tilde{L}^0}^2}{m_{\tilde{L}^{\pm}}^2}~,\label{eq:S}\\
\alpha T_{\tilde{L}} &=&  \frac{1}{16\pi m_{\rm W}^2 \sin^2\theta_{\rm W}}
\left[
m_{\tilde{L}^0}^2  +  m_{\tilde{L}^{\pm}}^2  +  \frac{2m_{\tilde{L}^0}^2 m_{\tilde{L}^{\pm}}^2}{(m_{\tilde{L}^0}^2 - m_{\tilde{L}^{\pm}}^2)}\ln\frac{m_{\tilde{L}^{\pm}}^2}{m_{\tilde{L}^{0}}^2}
\right]~,\label{eq:T} \\
\alpha U_{\tilde{L}} &=&  \frac{1}{12\pi}
\left[
-\frac{5m_{\tilde{L}^0}^4 - 22m_{\tilde{L}^0}^2m_{\tilde{L}^{\pm}}^2 + 5m_{\tilde{L}^{\pm}}^4}{3(m_{\tilde{L}^0}^2-m_{\tilde{L}^{\pm}}^2)^2} + 
\frac{m_{\tilde{L}^0}^6 -3m_{\tilde{L}^0}^4 m_{\tilde{L}^{\pm}}^2 -3 m_{\tilde{L}^{\pm}}^4m_{\tilde{L}^0}^2 + m_{\tilde{L}^{\pm}}^6}{(m_{\tilde{L}^0}^2 - m_{\tilde{L}^{\pm}}^2)^3}
\ln\frac{m_{\tilde{L}^0}^2}{m_{\tilde{L}^{\pm}}^2}
\right]~,\label{eq:U}
\end{eqnarray}
where $\alpha$ is the fine-structure constant.
Of particular concern are the $T$ and $U$ corrections since they measure the amount of weak isospin breaking, here represented by the mass 
splitting $\Delta m_{\tilde{L}}  \equiv m_{\tilde{L}^{\pm}} - m_{\tilde{L}^0}$.
\begin{figure}[!htb!]
  \includegraphics[width=0.4\textwidth]{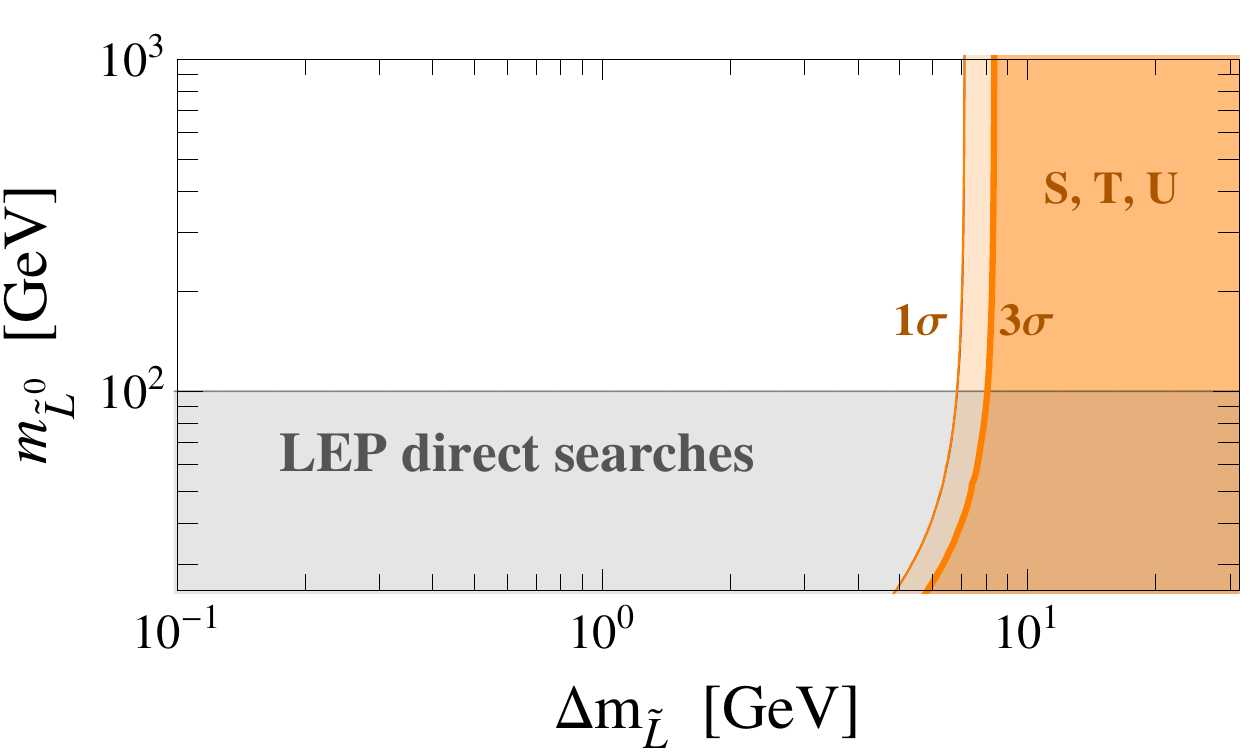}
 \caption{\textit{
 LEP bound on the mass spectrum of the scalar doublet $\tilde{L}=(\tilde{L}^0,\tilde{L}^-)^{\rm T}$. The gray region represents an indicative lower bound of the same order of the one set in the context of direct slepton searches~\cite{LEPIISUSY}. The orange regions represent the 1- and 3-$\sigma$ exclusion limits obtained 
 including eqs.~(\ref{eq:S})--(\ref{eq:U}) in the fit of LEP-I and LEP-II data~\cite{Falkowski:2013dza}.
  }}\label{fig:LEP}
\end{figure}
In fig.~\ref{fig:LEP} we show the 1- and 3-$\sigma$ exclusion limits on the mass spectrum of the scalar doublet obtained from LEP-I and LEP-II data; mass splittings $\Delta m_{\tilde{L}} \gtrsim  \mathcal{O}(10\,{\rm GeV})$ are in tension with EW precision data.
Considering the Lagrangian in eq.~(\ref{eq:LagrangianLtilde}), the two components of the scalar doublet $\tilde{L}$ are degenerate in mass at the tree level.
However, EW loop corrections induce a mass splitting such that  the charged component turns out to be slightly heavier than the neutral one. 
Explicitly, we have~\cite{Cirelli:2005uq} $\Delta m_{\tilde{L}}  = (\alpha_{\rm W}/4\pi)m_{\tilde{L}^0}\sin^2\theta_{\rm W}f\left(m_{\rm Z}/m_{\tilde{L}^0}\right)$ where, omitting the UV divergent part, $f(r)=-r\left[2r^3\ln r
+(r^2 - 4)^{3/2}\ln A
\right]/4$ with $A\equiv (r^2-2-r\sqrt{r^2-4})/2$. This would correspond to a mass 
splitting $\Delta m_{\tilde{L}} \sim 250$ MeV for $m_{\tilde{L}^0} = 150$ GeV. Thanks to this mass splitting, the charged component of $\tilde{L}$ is not stable;
the dominant decay channels are $\tilde{L}^{\pm} \to \tilde{L}^{0}W^{\pm\,*}\to \tilde{L}^{0}\pi^{\pm}, \tilde{L}^{0}e^{\pm}\nu_e, \tilde{L}^{0}\mu^{\pm}\nu_{\mu}$.
The first one turns out to be dominant~\cite{Cirelli:2005uq,FileviezPerez:2008bj}, with a typical decay length $c\tau_{\pi^{\pm}} \sim \mathcal{O}(10\,{\rm cm})$
if $\Delta m_{\tilde{L}}$ is of the order of  a few hundreds MeV's.

\item
EW Drell-Yan processes have three potential signatures: mono-X (where X can be a jet, a photon or an EW gauge boson), disappearing tracks and $W^{\pm}$ emission with subsequent lepton decay.
\begin{figure}[!htb!]
  \includegraphics[width=0.6\textwidth]{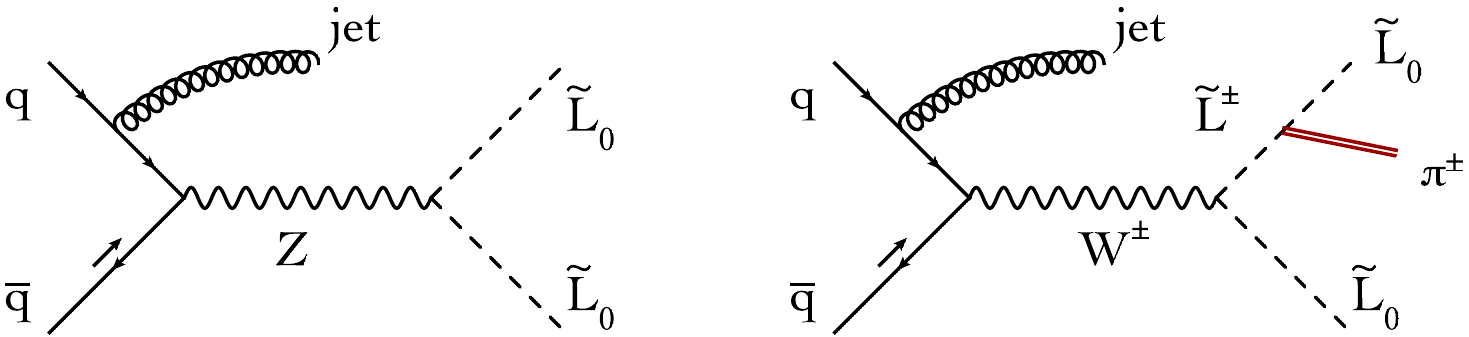}
 \caption{\textit{
 Representative Feynman diagrams (at the parton level) describing a typical mono-jet event with one hard jet and missing transverse energy in the final state  (left panel), and with 
  the additional presence of a disappearing track due to $\tilde{L}^{\pm}\to \pi^{\pm}\tilde{L}_0$ decay (right panel).
 }}\label{fig:MonoJet}
\end{figure}
As far as mono-jet analyses are concerned, mono-jet searches have been carried out both at the Tevatron~\cite{Abazov:2003gp} and 
the LHC~\cite{ATLAS:2012ky,Chatrchyan:2012me} in the context of large extra dimensions and DM via effective contact operators.
Similar studies have been proposed in the context of simplified supersymmetric model~\cite{Low:2014cba}
and minimal DM~\cite{Cirelli:2014dsa}.
Even if these analyses cannot be directly applied to our setup, they show that values $m_{\tilde{L}^0}$ of the order of  a few hundreds GeV's are beneath the reach of high-luminosity LHC with $\sqrt{s}= 14$ TeV.

As previously mentioned, the pions, electrons and muons produced via two- and three-body 
decays
$\tilde{L}^{\pm} \to \tilde{L}^{0}\pi^{\pm}, \tilde{L}^{0}e^{\pm}\nu_e, \tilde{L}^{0}\mu^{\pm}\nu_{\mu}$
are extremely soft (since $\Delta m_{\tilde{L}}\sim {\rm few}\,100$ MeV) and, as a consequence, invisible to the tracking system.
Therefore, analysis based on $W^{\pm}$ emission with leptons in the final state  (unless $\Delta m_{\tilde{L}}\gtrsim 10$ GeV, unrealistic in our minimal scenario) 
are disfavored.
On the contrary, the footprint of the soft decays $\tilde{L}^{\pm} \to \tilde{L}^{0}\pi^{\pm}, \tilde{L}^{0}e^{\pm}\nu_e, \tilde{L}^{0}\mu^{\pm}\nu_{\mu}$
is represented by the presence of a disappearing $\mathcal{O}(10\,{\rm cm})$ charged track.
At the LHC, ATLAS searches 
of disappearing tracks signatures in the context of simplified supersymmetric models with a chargino nearly mass-degenerate with the lightest neutralino
place a 95\% C.L. exclusion limit for $M_{\chi^{\pm}} < 270$ GeV with luminosity $L=20.3$ fb$^{-1}$~\cite{Aad:2013yna}.
Similar studies with simplified supersymmetric spectra~\cite{Low:2014cba}
and minimal DM~\cite{Cirelli:2014dsa} show that this bound can be further raised considering a center of mass energy of $\sqrt{s} = 14$ TeV
and $L=3$ ab$^{-1}$.

\end{enumerate}

\subsection{Two Higgs doublets}\label{sec:Higgses}

 The model features two Higgs doublets, $H^{u}$ ($Y = -1$) and $H^d$ ($Y = 1$).
The most general potential is~\cite{Branco:2011iw}
\begin{eqnarray}\label{eq:HiggsPotential}
V(H^u, H^d) &=& m_{u}^2 H^{u\,*}_{\alpha}H^{u}_{\alpha} + m_{d}^2 H^{d\,*}_{\alpha}H^{d}_{\alpha} 
+ \frac{\lambda_u}{2}\left(H^{u\,*}_{\alpha}H^{u}_{\alpha}\right)^2 + 
 \frac{\lambda_d}{2}\left(H^{d\,*}_{\alpha}H^{d}_{\alpha}\right)^2 + \lambda_3  H^{u\,*}_{\alpha}H^{u}_{\alpha}H^{d\,*}_{\beta}H^{d}_{\beta}
 + \lambda_4  H^{d\,*}_{\alpha}H^{u}_{\alpha}H^{u\,*}_{\beta}H^{d}_{\beta}\nonumber \\
 &+& \left[
 B_{\mu} \epsilon_{\alpha\beta}H^d_{\alpha} H^u_{\beta} +  \frac{\lambda_5}{2} (\epsilon_{\alpha\beta}H^d_{\alpha} H^u_{\beta})^2 -
  \lambda_6 (\epsilon_{\alpha\beta}H^d_{\alpha} H^u_{\beta})H^{u\,\dag}H^{u} -  \lambda_7 (\epsilon_{\alpha\beta}H^d_{\alpha} H^u_{\beta})H^{d\,\dag}H^{d}
 + h.c. 
 \right]~.
\end{eqnarray}
Referring to eqs.~(\ref{eq:Hup})--(\ref{eq:Hdown}), the two Higgs doublets are characterized by the usual Goldstone bosons $G^{\pm}$, $G^0$ that gives the longitudinal component of 
the $W^{\pm}$ and $Z$ gauge bosons, two CP-even scalars $h_0$ and $H^0$, one CP-even scalar $A^0$ and a charged Higgs $H^{\pm}$. 
In full generality, the $SU(2)_L$ breaking vacuum expectation values of the
 two doublets are $\langle H^u \rangle = (v\cos\beta/\sqrt{2}, 0)^{\rm T}$, $\langle H^d \rangle = (0, v\sin\beta/\sqrt{2})^{\rm T}$. 
 The two Higgs doublet model described by the potential in eq.~(\ref{eq:HiggsPotential}) has been studied in ref.~\cite{Altmannshofer:2012ar}. 
 Here, we just notice that our model is compatible with a simplified version of eq.~(\ref{eq:HiggsPotential}) in which a discrete $Z_2$ symmetry is enforced.
 Imposing the transformation rules $\Phi_1 \to - \Phi_1$, $\Phi_2 \to \Phi_2$, $\Phi_{\rm NP}\to -\Phi_{\rm NP}$
 it follows that $B_{\mu} = \lambda_6 = \lambda_7=0$. Moreover, since now $\langle H^u \rangle $ is forced to be zero, the only  
 $SU(2)_L$ breaking vacuum expectation value is $\langle H^d \rangle$. 
 In this setup the Higgs sector recovers the so-called Type I two-Higgs doublet model~\cite{Branco:2011iw} where 
 $H^d$---coupled to leptons, down- and (via $i\sigma_2H^{d\,*}$) up-type quarks---plays the role of the SM Higgs
while the second doublet $H^u$ is inert. 
The model predicts deviations of the tree level Higgs couplings to gauge bosons and fermions ($g_{hVV}$, $g_{hf\bar{f}}$) with respect to 
their SM values ($g_{hVV}^{\rm SM}$, $g_{hf\bar{f}}^{\rm SM}$). In particular we have
\begin{equation}
g_{hVV} = \cos\alpha~g_{hVV}^{\rm SM}~,~~~~g_{hf\bar{f}} = \cos\alpha~g_{hf\bar{f}}^{\rm SM}~.
\end{equation}
 
 The Higgs sector of the model offers a rich phenomenology that can be tested at the LHC
considering both the existence of new additional degrees of freedom and deviation from the SM Higgs couplings.

\subsection{Dark matter}
\label{sec:dm}

Does the model contain a  DM candidate?
The simplest possibility is to look for a viable DM candidate in the inert Higgs doublet as discussed  in the context of the so-called Inert Doublet Model~\cite{LopezHonorez:2006gr}. In this case DM can be identified either with $H^0$ or $A^0$, with a mass around $100$ GeV~\cite{Honorez:2010re}.

There also exists another  possibility, related to the presence in the model of the fermionic 
Higgs partner $\tilde{h}^d = (\tilde{h}^+_d,\tilde{h}^0_d)_{L}^{\rm T}$ (see eq.~(\ref{eq:Chiral2})).  
These Higgsinos are stable because the potentially dangerous decay channel arising---after EW symmetry breaking, and assuming the inert condition $\langle H^u \rangle = 0$---form the effective operator 
$\epsilon_{\alpha\beta}\epsilon_{\alpha'\beta'} (L_{\alpha}\cdot \tilde{h}^d_{\beta}H_{\alpha'}^uH_{\beta'}^{d} 
- L_{\alpha}\cdot \tilde{h}^d_{\alpha'}H_{\beta}^dH_{\beta'}^u + h.c.)/M^2$ in eq.~(\ref{eq:EffL}) 
can be kinematically closed by appropriately  choosing the mass spectrum  of the inert doublet. 

As already mentioned in section~\ref{sec:OneGeneration}, it is necessary to introduce an extra $SU(2)_L$ doublet with hypercharge $Y=-1$, $\tilde{h}^u$ hereafter,  in 
order to cancel the gauge anomaly.
The state $\tilde{h}^u$, together with $\tilde{h}^d$, can provide a good DM candidate. 
To show it, we 
explicitly introduce the following gauge invariant soft mass term
\begin{equation}\label{eq:DMmassterm}
\mathcal{L}_{\rm soft} = -\mu\,\epsilon_{\alpha\beta}\,\tilde{h}^{u}_{\alpha}\cdot  \tilde{h}^d_{\beta} + h.c. 
=-\mu\,\tilde{h}_u^{0} \cdot \tilde{h}_d^{0} +\mu\,\tilde{h}_u^{-}\cdot \tilde{h}_d^{+} + h.c. = 
-\frac{1}{2}(\tilde{h}_u^0, \tilde{h}_d^0)
\left(
\begin{array}{cc}
 0 & \mu   \\
 \mu  &  0  
\end{array}
\right)\left(
\begin{array}{c}
  \tilde{h}^0_u   \\
  \tilde{h}^0_d 
\end{array}
\right)+\mu\,\tilde{h}_u^{-}\cdot \tilde{h}_d^{+} + h.c.~,
\end{equation}
where, for reason of clarity, we used the following notation
\begin{equation}
\tilde{h}^u =
\left(
\begin{array}{c}
  \tilde{h}^0_u   \\
  \tilde{h}^-_u 
\end{array}
\right)_L~,~~~
\tilde{h}^d =
\left(
\begin{array}{c}
  \tilde{h}^+_d   \\
  \tilde{h}^0_d 
\end{array}
\right)_L~.
\end{equation}
The mass term in eq.~(\ref{eq:DMmassterm}) defines a degenerate spectrum of one charged (Dirac) fermion $\Psi \equiv (\tilde{h}^+_d, (\tilde{h}_u^-)^{\rm C})^{\rm T}$
plus two neutral fields
\begin{equation}\label{eq:MassHiggsinoEigenstates}
\chi_1 \equiv \frac{i}{\sqrt{2}}(\tilde{h}_u^0 - \tilde{h}_d^0)~,~~~~\chi_2 \equiv \frac{1}{\sqrt{2}}(\tilde{h}_u^0 + \tilde{h}_d^0)~,
\end{equation} 
obtained diagonalizing the mass matrix in the neutral sector by means of the unitary transformation $U\equiv
\left(
\begin{array}{cc}
i  & 1    \\
 -i &   1
\end{array}
\right)/\sqrt{2}
$. 
 The gauge coupling to the $Z$ boson 
\begin{equation}\label{eq:Zexchange}
\mathcal{L}_Z = -\frac{g}{2\cos\theta_W} \left(\bar{\tilde{h}}_u^0, \bar{\tilde{h}}_d^0\right)
\bar{\sigma}^{\mu}
\left(
\begin{array}{cc}
 1 & 0    \\
 0 &   -1
\end{array}
\right)Z_{\mu}
\left(
\begin{array}{c}
  \tilde{h}_u^0  \\
   \tilde{h}_d^0
\end{array}
\right) = 
-\frac{ig}{2\cos\theta_W}
\left(\bar{\chi}_1, \bar{\chi}_2\right)
\bar{\sigma}^{\mu}
\left(
\begin{array}{cc}
 0 & 1    \\
 1 & 0
\end{array}
\right)Z_{\mu}
\left(
\begin{array}{c}
  \chi_1  \\
  \chi_2
\end{array}
\right)\,,
\end{equation}
is non-diagonal in the mass eigenstates.

The setup outlined by these equations closely resembles 
the typical scenario of pure Higgsino DM, in which we have four co-annihilating degenerate states $(\chi_1, \chi_2, \tilde{h}^+_d,\tilde{h}^-_u)$.
The dominant annihilation channels are into EW gauge bosons, whereas $t$-channel annihilation processes mediated by $N$ and $\phi$ are suppressed 
by the large value of $M$. The thermally averaged effective annihilation cross-section times relative velocity is given by~\cite{ArkaniHamed:2006mb}
\begin{equation}\label{eq:RelicCrossSection}
\langle \sigma_{\rm eff} v\rangle = \frac{g^4}{512 \pi \mu^2}\left(21 + 3\tan^2\theta_W + 11\tan^4\theta_W \right)~,
\end{equation}
and, consequently, the relic abundance is set by 
\be
\Omega_{\tilde{h}} h^2 \approx 0.1 \left(\frac{\mu}{1~{\rm TeV}}\right)^2\,,
\ee
 to be compared with the observed value 
$\Omega_{\rm DM}h^2 = 0.1199\pm 0.0027$ ($68\%$ C.L.)~\cite{planck}. Therefore, in order to reproduced the totality of the observed DM density, one needs $\mu \sim \mathcal{O}(1\,{\rm TeV})$.\footnote{This conclusion should only be taken as indicative. Smaller values of $\mu$, for instance, are compatible with 
a cosmological scenario in which the decay of heavy gravitinos when DM is already out of equilibrium 
 increases its relic abundance~\cite{ArkaniHamed:2004yi}.} 
Referring to what mentioned at the beginning of this section, 
smaller values of $\mu$---and, as a consequence, smaller values of $\Omega_{\tilde{h}} h^2$---are consistent with a mixed scenario 
 in which  DM is made of two components, namely the lightest Higgsino state discussed here and the neutral component (either $H^0$ or $A^0$) of the inert Higgs doublet.
 
Bounds 
 imposed by direct detection experiments~\cite{Akerib:2013tjd}  constrain the proprieties of the Higgsino DM  candidate. 
 Referring to the previous discussion, 
 the most dangerous aspect that needs to be addressed is related to the necessity of removing the degeneracy
 between the states $(\chi_1, \chi_2, \tilde{h}^+_d,\tilde{h}^-_u)$. Two small (that is much smaller than $\mu$) mass splittings---a first one, $\Delta_+$, between the charged and the neutrals components and 
 a second one, $\Delta_0$,  between the two neutral components---do not affect the order of magnitude of the cross-section in eq.~(\ref{eq:RelicCrossSection})
 but, in addition to properly define the actual lightest particle inside the multiplet, play a fundamental role for direct detection.
 The $Z$ boson exchange in eq.~(\ref{eq:Zexchange}) leads to a large spin-independent DM-nucleon cross-section, already severely constrained by direct detection
 experiments. 
However, if the mass splitting is larger than the typical $\mathcal{O}(100\,{\rm keV})$ momentum transfer in DM-nucleon scatterings these constraints can be 
avoided since DM can not scatter elastically~\cite{Hall:1997ah}.
In the MSSM tree-level mixing with the gauginos and loop corrections introduce both $\Delta_+$ and $\Delta_0$ \cite{Giudice:1995np}. 
In our scenario, on the contrary, the only relevant corrections come from the EW gauge loops, and only $\Delta_+$ can be generated \cite{Masip:2005fv}.
Accordingly,  the Higgsino realization of the DM setup outlined in this section can be realized 
only by SUSY-promoting the $SU(2)_L$ gauge interactions, thus including EW gauginos.

\section{Conclusions}\label{sec:Conclusions}

In this paper we have shown that the idea of naturalness is still a useful heuristic principle in model building beyond the SM whenever a new mass threshold $M$
 above the EW scale is introduced. As a relevant example, we have discussed the generation of neutrino masses in the context of the type-I seesaw.
 In this case the new physics threshold is the mass scale $M$ of right-handed neutrinos.
 In summary, the distinctive features of our approach are the following:
\begin{itemize}
 
\item[$\circ$] We have introduced SUSY in order to protect the Higgs mass from large quantum corrections proportional to $M^2$. 
 The novelty is that only the new physics sector needs to be super-symmetrized while the SM is left unchanged. This new way of thinking directly follows from the fact that the SM by itself is a natural theory.
 
\item[$\circ$] Considering the simplest realization of this idea, gauge interactions break SUSY at two loops. 
Because of this breaking, the Higgs mass receives quantum corrections proportional to $M^2$ but featuring an extra suppression $\mathcal{O}(\alpha_{\rm W}/4\pi)$ with respect to the results commonly quoted in the literature. Thanks to this factor, we have shown that it is possible to obtain a right-handed neutrino mass scale that is, at the same time, natural and large enough to give rise to baryogenesis (via leptogenesis).

\item[$\circ$] We have shown that the model obtained following this approach is characterized by distinctive new physics. 
In particular, we have discussed bounds and signatures related to the existence of extra scalar and fermionic degrees of freedom, and outlined the properties related to the presence of a second Higgs doublet. Distinguishing low-energy signatures consist of disappearing charged tracks and mono-jet signals, both related to the presence of new weakly coupled states. 
More generally, the model proposed in this paper is characterized by the absence of colored SUSY partners such as gluinos and squarks. 
For example, the discovery of top partners at the LHC run II, therefore, would weaken our scenario. 
If neither top partners or gluinos will be discovered, on the contrary, the approach to naturalness outlined in this paper
may play a guidance role for model builders.

\item[$\circ$] If larger values of $M$ are required, the SM itself must be made supersymmetric. This can be implemented in stages. In the model here considered, this process starts with the EW gauge sector; its supersymmetrization moves to three-loop order the large $M^2$ terms---which are thus further suppressed. At the next order, also SUSY partners for the fermions must be  included. This approach shows the MSSM under a different light, one in which SUSY is not a fundamental symmetry and naturalness is preserved  by means of a  partially realized SUSY in which, in first instance, there are no  superpartners for the strongly interacting states.
 
 \end{itemize}

\begin{acknowledgments}
We thank Sacha Davidson and Michele Frigerio for discussions.
The work of A.U. is supported by the ERC Advanced Grant n.\ 267985, ``Electroweak Symmetry Breaking, Flavour and Dark Matter: One Solution for Three Mysteries" (DaMeSyFla). M.F. thanks SISSA for the hospitality.
\end{acknowledgments}




\begin{thebibliography}{99}


\bibitem{nat} 
K. Wilson, Phys. Rev. D3 (1971) 1818; 
G. 't Hooft, {\it Recent Developments in Gauge Theories} in Proc. 1979 Carg\'ese Summer Institute (ed. G. 't Hooft e al., Plenum Press, NY 1980).

\bibitem{natgut}
  E.~Gildener,
  Phys.\ Rev.\ D {\bf 14}, 1667 (1976); 
  Phys.\ Lett.\ B {\bf 92}, 111 (1980);
  S.~Weinberg,
  Phys.\ Lett.\ B {\bf 82}, 387 (1979) .
  
\bibitem{Rothstein:2003mp} 
 For example, see:  I.~Z.~Rothstein,
  \hhref{hep-ph/0308266}.

\bibitem{natural} For more recent work, see:
  W.~A.~Bardeen,
  FERMILAB-CONF-95-391-T;
     K.~A.~Meissner and H.~Nicolai,
  Phys.\ Lett.\ B {\bf 660}, 260 (2008)
  [\hhref{arXiv:0710.2840}];
    R.~Foot, A.~Kobakhidze, K.~L.~McDonald and R.~R.~Volkas,
  Phys.\ Rev.\ D {\bf 77}, 035006 (2008)
  [\hhref{arXiv:0709.2750} [hep-ph]];
  M.~Shaposhnikov and D.~Zenhausern,
  Phys.\ Lett.\ B {\bf 671}, 162 (2009)
  [\hhref{arXiv:0809.3406}];
   F.~Bazzocchi and M.~Fabbrichesi,
   Phys.\ Rev.\ D {\bf 87}, 036001 (2013)
   [\hhref{arXiv:1212.5065}];
    M. Farina, D. Pappadopulo and A. Strumia, JHEP 1308 (2013) 022 [\hhref{arXiv:1303.7244} [hep-ph]]; 
    M. Fabbrichesi and S.T. Petcov, Eur. Phys. J. C74 (2014) 2774 [\hhref{arXiv:1304.4001} [hep-ph]]; 
    M.~Heikinheimo, A.~Racioppi, M.~Raidal, C.~Spethmann and K.~Tuominen, Mod.\ Phys.\ Lett.\ A {\bf 29}, 1450077 (2014) [\hhref{arXiv:1304.7006} [hep-ph]];
    R. Foot et al., Phys.Rev. D89 (2014) 115018 [\hhref{arXiv:1310.0223} [hep-ph]];  
    A. de Gouvea et al., Phys. Rev. D89 (2014) 115005 [\hhref{arXiv:1402.2658} [hep-ph]]. 
   
 \bibitem{SUSYcanc}
   E.~Witten,
  Nucl.\ Phys.\ B {\bf 188}, 513 (1981);
  S.~Dimopoulos and H.~Georgi,
  Nucl.\ Phys.\ B {\bf 193}, 150 (1981);
    N.~Sakai,
  Z.\ Phys.\ C {\bf 11}, 153 (1981);
    R.~K.~Kaul,
  Phys.\ Lett.\ B {\bf 109}, 19 (1982).
   
  \bibitem{seesaw}
P.~Minkowski,
\newblock Phys. Lett. {\bf B67}, 421 (1977);
%
M.~Gell-Mann, P.~Ramond and R.~Slansky,
\newblock (1979),
\newblock Print-80-0576 (CERN);
%
T.~Yanagida,
\newblock (KEK lectures, 1979),
\newblock ed. Sawada and Sugamoto (KEK, 1979);
%
R.~N. Mohapatra and G.~Senjanovic,
\newblock Phys. Rev. Lett. {\bf 44}, 91 (1980).

\bibitem{Heikinheimo:2013xua} 
  M.~Heikinheimo, A.~Racioppi, M.~Raidal, C.~Spethmann and K.~Tuominen,
  Nucl.\ Phys.\ B {\bf 876}, 201 (2013)
  [\hhref{arXiv:1305.4182} [hep-ph]].

\bibitem{Martin:1997ns} 
  S.~P.~Martin,
  Adv.\ Ser.\ Direct.\ High Energy Phys.\  {\bf 21}, 1 (2010)
  [\hhref{hep-ph/9709356}].
  
 
\bibitem{Wess:1973kz} 
  J.~Wess and B.~Zumino,
  Phys.\ Lett.\ B {\bf 49}, 52 (1974).
 
  
 
  
 \bibitem{Jurcisinova:2010zz} 
  E.~Jurcisinova and M.~Jurcisin,
  Phys.\ Lett.\ B {\bf 692}, 57 (2010).
  
  
\bibitem{Smirnov:2012gma} 
  V.~A.~Smirnov,
  Springer Tracts Mod.\ Phys.\  {\bf 250}, 1 (2012).

\bibitem{Frampton:2002qc} 
  P.~H.~Frampton, S.~L.~Glashow and T.~Yanagida,
  Phys.\ Lett.\ B {\bf 548}, 119 (2002)
  [\hhref{hep-ph/0208157}].
  
  
 
\bibitem{Vissani:1997ys} 
  F.~Vissani,
  Phys.\ Rev.\ D {\bf 57}, 7027 (1998)
  [\hhref{hep-ph/9709409}].
  
\bibitem{Pilaftsis:2003gt} 
  A.~Pilaftsis and T.~E.~J.~Underwood,
  Nucl.\ Phys.\ B {\bf 692}, 303 (2004)
  [\hhref{hep-ph/0309342}].
  
\bibitem{Pilaftsis:2005rv} 
  A.~Pilaftsis and T.~E.~J.~Underwood,
  Phys.\ Rev.\ D {\bf 72}, 113001 (2005)
  [\hhref{hep-ph/0506107}].
  
\bibitem{Dev:2014laa} 
  P.~S.~Bhupal Dev, P.~Millington, A.~Pilaftsis and D.~Teresi,
  Nucl.\ Phys.\ B {\bf 886}, 569 (2014)
  [\hhref{arXiv:1404.1003} [hep-ph]].
  
 
 
  
 
\bibitem{Davidson:2002qv} 
  G.~F.~Giudice, A.~Notari, M.~Raidal, A.~Riotto and A.~Strumia,
  Nucl.\ Phys.\ B {\bf 685}, 89 (2004)
  [\hhref{hep-ph/0310123}];
    S.~Davidson, E.~Nardi and Y.~Nir,
  Phys.\ Rept.\  {\bf 466}, 105 (2008)
  [\hhref{arXiv:0802.2962} [hep-ph]];
    P.~Di Bari,
  Contemp.\ Phys.\  {\bf 53}, no. 4, 315 (2012)
  [\hhref{arXiv:1206.3168} [hep-ph]].


 

\bibitem{Clarke:2015gwa} 
  J.~D.~Clarke, R.~Foot and R.~R.~Volkas,
  \hhref{arXiv:1502.01352} [hep-ph].


\bibitem{Kawasaki:1994af}
  M.~Kawasaki and T.~Moroi,
  Prog.\ Theor.\ Phys.\  {\bf 93} (1995) 879
  [\hhref{hep-ph/9403364}];
  R.~H.~Cyburt, J.~R.~Ellis, B.~D.~Fields and K.~A.~Olive,
  Phys.\ Rev.\ D {\bf 67}, 103521 (2003)
  [\hhref{astro-ph/0211258}];
    M.~Kawasaki, K.~Kohri, T.~Moroi and A.~Yotsuyanagi,
  Phys.\ Rev.\ D {\bf 78}, 065011 (2008)
  [\hhref{arXiv:0804.3745} [hep-ph]].

\bibitem{CMS:2013bda} 
  CMS Collaboration [CMS Collaboration],
  CMS-PAS-SUS-12-022.
 
  
    \bibitem{LEPIISUSY} 
For example, see: \href{http://lepsusy.web.cern.ch/lepsusy/www/sleptons_budapest01/sleptons_pub.html}{LEP2 SUSY Working Group}
  
\bibitem{Peskin:1990zt} 
  M.~E.~Peskin and T.~Takeuchi,
  Phys.\ Rev.\ Lett.\  {\bf 65}, 964 (1990);
  M.~Golden and L.~Randall,
  Nucl.\ Phys.\ B {\bf 361}, 3 (1991);
  B.~Holdom and J.~Terning,
  Phys.\ Lett.\ B {\bf 247}, 88 (1990).;
  M.~E.~Peskin and T.~Takeuchi,
  Phys.\ Rev.\ D {\bf 46}, 381 (1992);
  G.~Altarelli and R.~Barbieri,
  Phys.\ Lett.\ B {\bf 253}, 161 (1991);
  G.~Altarelli, R.~Barbieri and S.~Jadach,
  Nucl.\ Phys.\ B {\bf 369}, 3 (1992)
  [Erratum-ibid.\ B {\bf 376}, 444 (1992)].

\bibitem{Barbieri:2003pr} 
  R.~Barbieri, A.~Pomarol and R.~Rattazzi,
  Phys.\ Lett.\ B {\bf 591}, 141 (2004)
  [\hhref{hep-ph/0310285}];
  R.~Barbieri, A.~Pomarol, R.~Rattazzi and A.~Strumia,
  Nucl.\ Phys.\ B {\bf 703}, 127 (2004)
  [\hhref{hep-ph/0405040}].
  
\bibitem{Blum:2011fa} 
  K.~Blum, Y.~Hochberg and Y.~Nir,
  JHEP {\bf 1110}, 124 (2011)
  [\hhref{arXiv:1107.4350} [hep-ph]];
  H.~H.~Zhang, W.~B.~Yan and X.~S.~Li,
  Mod.\ Phys.\ Lett.\ A {\bf 23}, 637 (2008)
  [\hhref{hep-ph/0612059}];
  W.~C.~Huang and A.~Urbano,
  JHEP {\bf 1303}, 079 (2013)
  [\hhref{arXiv:1212.1399} [hep-ph]].
  
\bibitem{Falkowski:2013dza} 
  A.~Falkowski, F.~Riva and A.~Urbano,
  JHEP {\bf 1311}, 111 (2013)
  [\hhref{arXiv:1303.1812} [hep-ph]].
  
  

\bibitem{Cirelli:2005uq} 
  M.~Cirelli, N.~Fornengo and A.~Strumia,
  Nucl.\ Phys.\ B {\bf 753}, 178 (2006)
  [\hhref{hep-ph/0512090}].

\bibitem{FileviezPerez:2008bj} 
  P.~Fileviez Perez, H.~H.~Patel, M.~J.~Ramsey-Musolf and K.~Wang,
  Phys.\ Rev.\ D {\bf 79}, 055024 (2009)
  [\hhref{arXiv:0811.3957} [hep-ph]].
  
  
\bibitem{Abazov:2003gp} 
  V.~M.~Abazov {\it et al.}  [D0 Collaboration],
  Phys.\ Rev.\ Lett.\  {\bf 90}, 251802 (2003)
  [\hhref{hep-ex/0302014}];
  V.~M.~Abazov {\it et al.}  [D0 Collaboration],
  Phys.\ Rev.\ Lett.\  {\bf 101}, 011601 (2008)
  [\hhref{arXiv:0803.2137} [hep-ex]].

\bibitem{ATLAS:2012ky} 
  G.~Aad {\it et al.}  [ATLAS Collaboration],
  JHEP {\bf 1304}, 075 (2013)
  [\hhref{arXiv:1210.4491} [hep-ex]].
  
\bibitem{Chatrchyan:2012me} 
  S.~Chatrchyan {\it et al.}  [CMS Collaboration],
  JHEP {\bf 1209}, 094 (2012)
  [\hhref{arXiv:1206.5663} [hep-ex]].
  
\bibitem{Low:2014cba} 
  M.~Low and L.~T.~Wang,
  JHEP {\bf 1408}, 161 (2014)
  [\hhref{arXiv:1404.0682} [hep-ph]];
  G.~Grilli di Cortona,
  \hhref{arXiv:1412.5952} [hep-ph].
  
\bibitem{Cirelli:2014dsa} 
  M.~Cirelli, F.~Sala and M.~Taoso,
  JHEP {\bf 1410}, 033 (2014)
  [Erratum-ibid.\  {\bf 1501}, 041 (2015)]
  [\hhref{arXiv:1407.7058} [hep-ph]].
    
\bibitem{Aad:2013yna} 
  G.~Aad {\it et al.}  [ATLAS Collaboration],
  Phys.\ Rev.\ D {\bf 88}, no. 11, 112006 (2013)
  [\hhref{arXiv:1310.3675} [hep-ex]].
    
 
    
\bibitem{Branco:2011iw} 
 For a review, see, for example: G.~C.~Branco, P.~M.~Ferreira, L.~Lavoura, M.~N.~Rebelo, M.~Sher and J.~P.~Silva,
  Phys.\ Rept.\  {\bf 516}, 1 (2012)
  [\hhref{arXiv:1106.0034} [hep-ph]].
  
\bibitem{Altmannshofer:2012ar} 
  W.~Altmannshofer, S.~Gori and G.~D.~Kribs,
  Phys.\ Rev.\ D {\bf 86}, 115009 (2012)
  [\hhref{arXiv:1210.2465} [hep-ph]].
    
\bibitem{LopezHonorez:2006gr} 
  L.~Lopez Honorez, E.~Nezri, J.~F.~Oliver and M.~H.~G.~Tytgat,
  JCAP {\bf 0702}, 028 (2007)
  [\hhref{hep-ph/0612275}].
    
\bibitem{Honorez:2010re} 
  L.~Lopez Honorez and C.~E.~Yaguna,
  JHEP {\bf 1009}, 046 (2010)
  [\hhref{arXiv:1003.3125} [hep-ph]].
  
    
\bibitem{ArkaniHamed:2006mb} 
  N.~Arkani-Hamed, A.~Delgado and G.~F.~Giudice,
  Nucl.\ Phys.\ B {\bf 741}, 108 (2006)
  [\hhref{hep-ph/0601041}].

\bibitem{planck}
  P.~A.~R.~Ade {\it et al.}  [Planck Collaboration],
  \hhref{arXiv:1303.5076} [astro-ph.CO].
  
\bibitem{ArkaniHamed:2004yi} 
  N.~Arkani-Hamed, S.~Dimopoulos, G.~F.~Giudice and A.~Romanino,
  Nucl.\ Phys.\ B {\bf 709}, 3 (2005)
  [\hhref{hep-ph/0409232}].

\bibitem{Akerib:2013tjd} 
  D.~S.~Akerib {\it et al.}  [LUX Collaboration],
  Phys.\ Rev.\ Lett.\  {\bf 112}, 091303 (2014)
  [\hhref{arXiv:1310.8214} [astro-ph.CO]].

\bibitem{Hall:1997ah} 
  L.~J.~Hall, T.~Moroi and H.~Murayama,
  Phys.\ Lett.\ B {\bf 424}, 305 (1998)
  [\hhref{hep-ph/9712515}];
    D.~Tucker-Smith and N.~Weiner,
  Phys.\ Rev.\ D {\bf 64}, 043502 (2001)
  [\hhref{hep-ph/0101138}].

\bibitem{Giudice:1995np} 
  G.~F.~Giudice and A.~Pomarol,
  Phys.\ Lett.\ B {\bf 372}, 253 (1996)
  [\hhref{hep-ph/9512337}].

\bibitem{Masip:2005fv} 
  M.~Masip and I.~Mastromatteo,
  Phys.\ Rev.\ D {\bf 73}, 015007 (2006)
  [\hhref{hep-ph/0510311}].
  


\end{thebibliography}
\end{document}